%% file: flares_agn_ionising.tex
\documentclass[twocolumn]{openjournal}
\usepackage{natbib}
\usepackage{graphicx,amsmath,amssymb,amstext}
\usepackage{amsbsy,amsfonts,amsthm,color,xcolor}
\usepackage{lipsum}
\usepackage[colorlinks,linkcolor=blue,citecolor=blue,urlcolor=blue ]{hyperref}
\usepackage[utf8]{inputenc}
\usepackage{float}
\usepackage[caption=false]{subfig}
\usepackage{upgreek}

\newcommand{\qsosed}{\texttt{qsosed}}
\newcommand{\agnsed}{\texttt{agnsed}}

\newcommand{\eagle}{\mbox{EAGLE}}

\newcommand{\flares}{\mbox{FLARES}}

\newcommand{\Lbolagn}{L_{\rm bol,\bullet}/{\rm erg\ s^{-1}}}

\newcommand{\Lbol}{L_{\rm bol}/{\rm erg\ s^{-1}}}
\newcommand{\LHa}{L_{\rm H\alpha}/{\rm erg\ s^{-1}}}
\newcommand{\fcov}{f_{\rm cov}}
\newcommand{\fblr}{f_{\rm BLR}}
\newcommand{\tauagn}{\tau_{\rm H\alpha, \bullet}}

\newcommand{\Msun}{{\rm M_{\odot}}}

\newcommand{\Mbh}{M_{\bullet}/\Msun}

\defcitealias{lovell2020}{\flares\,I}

\newcommand\blfootnote[1]{%
  \begingroup
  \renewcommand\thefootnote{}\footnote{#1}%
  \addtocounter{footnote}{-1}%
  \endgroup
}

\begin{document}

\title{First Light and Reionization Epoch Simulations (FLARES) - XVIII: the ionising emissivities and hydrogen recombination line properties of early AGN\vspace{-3em}}

\author{Stephen M. Wilkins$^{1,2\dagger}$} 
\author{Aswin P. Vijayan$^{1}$} 
\author{Scott Hagen$^{3}$} %

\author{Joseph Caruana$^{2,4}$}
\author{Christopher J. Conselice$^{5}$}
\author{Chris Done$^{3}$} 
\author{Michaela Hirschmann$^{6}$}
\author{Dimitrios Irodotou$^{7}$} 
\author{Christopher C. Lovell$^{8}$} 
\author{Jorryt Matthee$^{9}$}
\author{Adèle Plat$^{6}$}
\author{William J. Roper$^{1}$} 
\author{Anthony J. Taylor$^{10}$}

\blfootnote{$^{\dagger}$ Corresponding author. Email: \href{mailto:s.wilkins@sussex.ac.uk}{s.wilkins@sussex.ac.uk}}

\affiliation{$^1$Astronomy Centre, University of Sussex, Falmer, Brighton BN1 9QH, UK}
\affiliation{$^2$Institute of Space Sciences and Astronomy, University of Malta, Msida MSD 2080, Malta}
\affiliation{$^3$Centre for Extragalactic Astronomy, Department of Physics, Durham University, South Road, Durham, DH1 3LE, UK}
\affiliation{$^4$Department of Physics, Faculty of Science, University of Malta, Msida MSD 2080, Malta}
\affiliation{$^5$Jodrell Bank Centre for Astrophysics, University of Manchester, Oxford Road, Manchester M13 9PL, UK}
\affiliation{$^6$Institute for Physics, EPFL, Observatoire de Sauverny, Chemin Pegasi 51, 1290 Versoix, Switzerland}
\affiliation{$^7$The Institute of Cancer Research, 123 Old Brompton Road, London SW7 3RP, UK}
\affiliation{$^8$Institute of Cosmology and Gravitation, University of Portsmouth, Burnaby Road, Portsmouth, PO1 3FX, UK}
\affiliation{$^9$Institute of Science and Technology Austria (ISTA), Am Campus 1, 3400 Klosterneuburg, Austria}
\affiliation{$^{10}$Department of Astronomy, The University of Texas at Austin, Austin, TX, USA}

\begin{abstract}

One of the most remarkable results from the \emph{James Webb Space Telescope} has been the discovery of a large population of compact sources exhibiting strong broad H$\alpha$ emission, typically interpreted to be low-luminosity broad-line (Type 1) active galactic nuclei (BLAGN). An important question is whether these observations are in tension with galaxy formation models, and if so how? While comparisons have been made using physical properties (i.e.~black hole mass and accretion rate) inferred from observations, these require the use of SED modelling assumptions, or locally inferred scaling relations, which may be unjustified, at least in the distant high-redshift Universe. In this work we take an alternative approach and forward model predictions from the First Light And Reionisation Epoch Simulations (FLARES) suite of cosmological hydrodynamical zoom simulations to predict the observable properties of BLAGN. We achieve this by first coupling \flares\ with the \qsosed\ model to predict the ionising photon luminosities of high-redshift ($z>5$) AGN. To model the observed broad H$\alpha$ emission we then assume a constant conversion factor and covering fraction, and the fraction of AGN that have observable broad-lines. With a reasonable choice of these parameters, \flares\ is able to reproduce observational constraints on the H$\alpha$ luminosity function and equivalent width distribution at $z=5$. 

\end{abstract}


\maketitle



\input sections/intro

\input sections/modelling

\input sections/results

\input sections/conclusion

\newpage

\section*{Author Contributions}

We list here the roles and contributions of the authors according to the Contributor Roles Taxonomy (CRediT)\footnote{\url{https://credit.niso.org/}}.
\textbf{Stephen M. Wilkins}: Conceptualization, Data curation, Methodology, Investigation, Formal Analysis, Visualization, Writing - original draft.
\textbf{Scott Hagen, Aswin P. Vijayan}: Conceptualization, Data curation, Methodology, Writing - original draft.
\textbf{Joseph Caruana, Christopher J. Conselice, Chris Done, Michaela Hirschmann, Dimitrios Irodotou, Christopher C. Lovell, Jorryt Matthee, Adèle Plat, William Roper, Anthony J. Taylor}: Writing - review \& editing.
    
\section*{Acknowledgements}

We thank the \eagle\, team for their efforts in developing the \eagle\, simulation code.  This work used the DiRAC@Durham facility managed by the Institute for Computational Cosmology on behalf of the STFC DiRAC HPC Facility (\url{www.dirac.ac.uk}). The equipment was funded by BEIS capital funding via STFC capital grants ST/K00042X/1, ST/P002293/1, ST/R002371/1 and ST/S002502/1, Durham University and STFC operations grant ST/R000832/1. DiRAC is part of the National e-Infrastructure. 

We also wish to acknowledge the following open source software packages used in the analysis: \textsc{Numpy} \citep{numpy}, \textsc{Scipy} \citep{scipy}, \textsc{Astropy} \citep{astropy:2013, astropy:2018, astropy:2022}, \textsc{Cmasher} \citep{cmasher}, and \textsc{Matplotlib} \citep{matplotlib}.

WJR, APV, and SMW acknowledge support from the Sussex Astronomy Centre STFC Consolidated Grant (ST/X001040/1). SH acknowledges support from the Science and Technologies Facilities Council (STFC) through the studentship grant ST/W507428/1. CD acknowledges support from STFC through grant ST/T000244/1. CJC acknowledges support from the ERC Advanced Investigator Grant EPOCHS (788113). CCL acknowledges support from a Dennis Sciama fellowship funded by the University of Portsmouth for the Institute of Cosmology and Gravitation. 

\section*{Data Availability Statement}

The data associated with the paper will be made publicly available at \href{https://flaresimulations.github.io}{https://flaresimulations.github.io} on the acceptance of the manuscript.



\bibliographystyle{mnras}
\bibliography{flares_agn_ionising, flares} 




\end{document}

%% file: sections/intro.tex
\section{Introduction}\label{sec:intro}

Super-massive black holes (SMBHs) are well established to play an important role in the formation and evolution of galaxies, particularly in the most massive galaxies \citep[e.g.][]{DiMatteo2005, Bower2006, Croton2006}. Highly accreting SMBHs, i.e. active galactic nuclei (AGN), are also amongst the most luminous objects in the Universe providing insights into e.g. structure formation, reionisation, etc.. However, a complete theoretical understanding of SMBHs/AGN is far from being achieved, with their origin, dynamics, growth, and impact on their hosts, still the subject of considerable study \citep[see][for a recent overview of SMBH modelling in cosmological simulations]{Habouzit2022a, Habouzit2022b}. Key to understanding these processes, and in particular their origin, may be observations of the earliest SMBHs, identified in the distant high-redshift ($z\simeq 6$) Universe. 



Bright examples (i.e. quasars) have been discovered by wide area imaging surveys combined with spectroscopic follow-up for more than 20 years \citep[][]{Fan2001, Mortlock2011, Willott2010, Jiang2016, Banados2018}. However, complementing these samples of rare, bright AGN, the \emph{James Webb Space Telescope (JWST)} has discovered samples of lower-luminosity ($L_{\rm bol}<10^{46}\ {\rm erg\ s^{-1}}$) and low-mass ($M_{\bullet}<10^{7}\ {\rm M_{\odot}}$) AGN. Samples have been identified spectroscopically, via broad Hydrogen lines, i.e. Broad Line (BL) or Type 1 AGN \citep{Matthee23, Larson2023, Kocevski2023, Kokorev23, Harikane2023, Kocevski2024, Maiolino24b, Taylor2024, Greene2024, Napolitano2025}, Narrow Line (NL) or Type 2 AGN \citep[e.g.]{Scholtz2023}, or photometrically \citep[e.g][]{Juodzbalis23, Kocevski24, Akins24, Labbe25}, albeit with less confidence. The innovation here is the resolution, sensitivity, but especially wavelength range, of \emph{Webb}'s capabilities. While samples are still relatively small they are rapidly growing in size (and diversity) but also in the breadth of spectroscopic and photometric constraints. For example, for increasing numbers of AGN we now not only have constraints on the H$\alpha$ emission but also other emission lines \citep[e.g][]{Brooks2024, Naidu2025, Rusakov2025} and continuum spectroscopy and/or photometry across the rest-frame UV and optical \citep[e.g][]{Ji2025, Naidu2025, Rusakov2025}, and beyond \citep[e.g][]{Gloudemans2025, RojasRuiz2025}. Many of the confirmed high-redshift BLAGN manifest as the so-called Little Red Dots (LRDs); however, the overlap between BLAGN and LRDs, i.e. the fraction of LRDs that are BLAGN, and vice-versa, remains debated, since some LRDs could also have a non-AGN origin \citep[e.g][]{Wang2024a, Setton2024, Wang2024b}. Moreover, even where there is the clear detection of broad H$\alpha$ emission, there also remains the possibility that this is not due to an AGN but some other mechanism, such as an outflow or high stellar velocity dispersion \citep[e.g][]{Baggen2024}.

Interpreting these observations in the context of modern cosmological galaxy formation models is particularly challenging for three reasons: (1) there is a vast difference in scale between the immediate environments of SMBHs ($<1$ pc) and cosmological volumes ($>10$ Mpc), this demands that SMBH physics in cosmological simulations is implemented using highly uncertain sub-grid models that must account for SMBH formation, dynamics, growth, and feedback; (2) most simulations lack the combination of volume and resolution to simulate sufficient numbers of bright AGN at the requisite resolution and (3) models do not \emph{ab initio} predict the observable quantities that are now accessible. The second challenge can be overcome by large periodic boxes, typically limited to high-redshift systems \citep[e.g][]{Feng2016, DiMatteo2017, Wilkins2017, Huang2018, Bird2022, Ni2022}, or by using a re-simulation approach, such as that employed by the First Light And Reionisation Epoch Simulations \citep[FLARES][]{FLARES-I, FLARES-II}. The third challenge arises because simulations \emph{ab initio} predict \emph{physical} properties, including black hole masses and accretion rates (or bolometric luminosities) and not directly observable quantities like $H\alpha$ luminosities. To make comparisons between models and observations the typical approach is to assume either a constant, or luminosity dependent, bolometric correction converting the observed H$\alpha$ luminosity to a bolometric luminosity, and thus accretion rate (assuming some radiative efficiency). However, the shape of the disc spectrum, and thus the bolometric correction, is predicted to depend on the properties of the SMBH and its disc, including its mass, accretion rate, spin, and the inclination of the disc \citep[e.g.][]{KD18, Hagen2023}. Assuming a single, or even luminosity dependent, bolometric correction will introduce a source of bias. Similarly, SMBH masses are inferred from the full-width half-maximum (FWHM) and luminosity of the H$\alpha$ line, typically assuming local empirical calibrations \citep[e.g.][]{Greene2005, Reines2013}. SMBH mass functions inferred solely from BLAGN samples will also be incomplete, missing obscured (i.e. Type 2) and low-luminosity, and low Eddington-ratio AGN, further complicating their interpretation.

In this work, we take the opposite approach to comparing observed and modelled samples, and instead forward model our simulated SMBHs to predict Hydrogen recombination line luminosities and equivalent widths for AGN. Here we use predictions from phase 1 of the First Light And Reionisation Simulations \citep[\flares][]{FLARES-I, FLARES-II} project, whose AGN demographics were previously published in \citet{FLARES-XIV}. To achieve this, we couple the \flares\ predictions with the \qsosed\ \citep{KD18} disc emission model, taking account of the SMBH mass and accretion rates, predictions from simple photoionisation modelling, and a simple model to account for obscuration and geometrical effects.

This article is organised as follows: In Section \ref{sec:modeling} we describe our modelling approach including introduction \flares\ (\S\ref{sec:modeling:flares}) and our modelling of the AGN emission (\S\ref{sec:modeling:agnemission}). In Section \ref{sec:results} we present our results including predictions for the ionising (\S\ref{sec:results:ionising}) and H$\alpha$ properties (\S\ref{sec:results:ha}) of AGN in \flares. In Section \S\ref{sec:discussion} we present a discussion of our results, including the limitations. In Section \ref{sec:conclusions} we present our conclusions.

%% file: sections/modelling.tex
\section{Modelling}\label{sec:modeling}

\subsection{FLARES}\label{sec:modeling:flares}

\flares\ is a suite of 40 hydrodynamical re-simulations of spherical regions of size $\rm 14\,cMpc\,$$h^{-1}$ drawn from a large (3.2 cGpc)$^3$ low-resolution dark matter only simulation \citep{CEAGLE}. The \flares\ regions encompass a wide range of environments, $\delta_{14}(z=5) = -0.4\to 1.0$, but include greater representation at extreme over-densities. This approach makes it possible to simulate many more massive, rare galaxies than possible using a periodic simulation with the same computational resources. This makes \flares\ ideally suited to studying SMBHs, and particularly AGN dominated systems, since they preferentially occur in the most massive galaxies. 

\flares\ adopts the AGNdT9 variant of the \eagle\ \citep{schaye_eagle_2015, crain_eagle_2015} physics model and a Planck year 1 cosmology \citep[$\rm \Omega_{M} = 0.307, \; \Omega_{\Lambda} = 0.693, \; $h$ \; = 0.6777$,][]{planck_collaboration_2014}. For details of the underlying physics model see \citet{schaye_eagle_2015} and \citet{crain_eagle_2015} and for details on \flares\ see \citet{FLARES-I, FLARES-II}.

\subsubsection{SMBHs in FLARES}

The physical properties of SMBHs in \flares\ are presented in \citet{FLARES-XIV} and we direct the reader to that paper, and the original EAGLE articles \citep{schaye_eagle_2015, crain_eagle_2015} for a full description of the model and predictions of key physical properties from FLARES. In short, halos are seeded with a $M_{\rm \bullet,\ seed} = 10^{5}\ h^{-1} {\rm M_{\odot}}$ SMBH when the halo reaches $M_{\rm h} = 10^{10}\ h^{-1} {\rm M_{\odot}}$. SMBHs are then able to grow through accretion and mergers. Growth via accretion follows a modified Bondi-Hoyle prescription but is capped at $1/h$ times the Eddington limit ($\dot{M}_{\rm Edd}$). SMBHs return energy via stochastic thermal feedback, stochastically heating surrounding particles by $\Delta T_{\rm AGN}$ ($=10^{9}\ {\rm K}$ in FLARES).   

As in \citet{FLARES-XIV}, to avoid numerical noise, we use the SMBH accretion rate averaged over 10 Myr when computing accretion rates, the bolometric luminosity, and thus other observable quantities. This is a significant limitation of this analysis since observed AGN vary on much shorter timescales. This could cause AGN luminosities to scatter, for example changing the shape of the luminosity function.

\begin{figure}
    \centering
    \includegraphics[width=1\columnwidth]{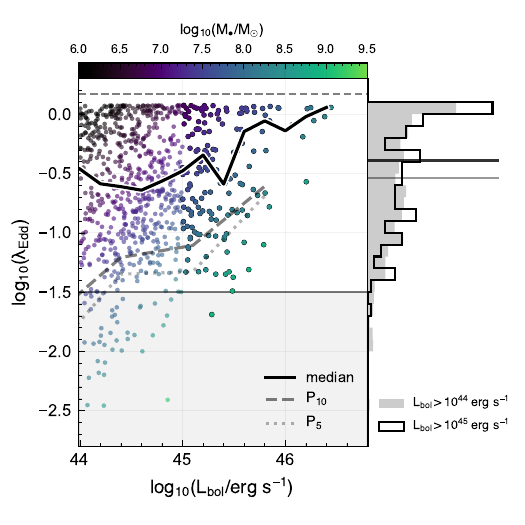}
    \caption{The relationship between the bolometric luminosity and Eddington ratio $\lambda_{\rm Edd}$ of SMBHs predicted by FLARES at $z=5$. Individual SMBHs are colour-coded by the mass. The solid horizontal line denotes the limit of the \qsosed\ modelling and the horizontal dashed line denotes the maximum Eddington ratio in FLARES/EAGLE ($\lambda_{\rm Edd}=h^{-1}$). The outlined curves show the binned median and 5$^{\rm th}$, and 10$^{\rm th}$ percentiles. The right-hand panel shows a histogram of the $\lambda_{\rm Edd}$ for $\Lbol>10^{44}$ (faint, filled histogram) and $\Lbol>10^{45}$ (dark, un-filled) with the faint and dark horizontal lines denoting the $\Lbol>10^{44}$ ($>10^{45}$) median values respectively.}
    \label{fig:modelling:bolometric_luminosity-eddington}
\end{figure}

In Figure \ref{fig:modelling:bolometric_luminosity-eddington} we show the Eddington ratio $\lambda_{\rm Edd}=\dot{M}_{\bullet}/\dot{M}_{\bullet,\rm Edd}$ as a function of the bolometric luminosity for all SMBHs with $\Lbol>10^{44}$ at $z=5$, which we use as a limit for our most liberal selection. Almost all SMBHs with $\Lbol>10^{44}$ have masses $\Mbh>10^{6}$ (c.f. the seed mass $=10^{5}\ h^{-1} {\rm M_\odot}$); the exception being a handful ($\sim 10$). These SMBHs have Eddington ratios $0.003\lesssim\lambda_{\rm Edd}\lesssim 1$, though with relatively few at $\lambda_{\rm Edd}<0.03$. This threshold is important because in this work we utilise the \qsosed\ \citep{KD18} disc model which is only robust at $\lambda_{\rm Edd}>0.03$. The median Eddington ratio across the entire sample is $\lambda_{\rm Edd}\approx 0.3$, though it increases with the bolometric luminosity, reaching $\lambda_{\rm Edd}\approx 1$ in our most luminous ($\Lbol\gtrsim 10^{46}$) SMBHs. In addition to our \emph{liberal} selection ($\Lbol>10^{44}$) we also define a \emph{conservative} sample with $\Lbol>10^{45}$. All but a single SMBH in the conservative sample have $\lambda_{\rm Edd}> 0.3$ with almost all also having $\Mbh>10^{7}$. The $\lambda_{\rm Edd}$ distribution of the two samples is similar in shape, but the conservative sample has a slightly higher average and more extreme high values.

\begin{figure}
    \centering
    \includegraphics[width=1\columnwidth]{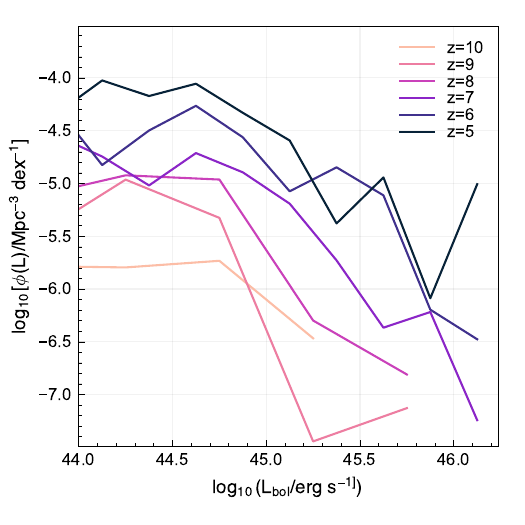}
    \caption{The bolometric luminosity function at $5\le z\le10$ of SMBHs predicted by FLARES.}
    \label{fig:modelling:bolometric_luminosity_function}
\end{figure}

Figure \ref{fig:modelling:bolometric_luminosity_function} shows the SMBH bolometric function  from $z=5\to 10$. Above redshift $z=5$ the normalisation of the luminosity function drops significantly ($\approx 1$ dex, $z=5\to 8$), with a corresponding increase in statistical uncertainty. For this reason we focus this analysis on $z=5$, but do extend our predictions to $z=6,7$ in \S\ref{sec:results:ha:half}.

\subsection{AGN Emission}\label{sec:modeling:agnemission}

In this work we adopt a simplified approach to modelling the broad-line H$\alpha$ emission from the accreting SMBH. In short, we associate every SMBH with an accretion flow and corresponding spectral energy distribution using the \qsosed\ model of \citet{KD18} (\S\ref{sec:modeling:agnemission:qsosed}) before applying a conversion, based on simple photoionisation modelling, to calculate the intrinsic H$\alpha$ properties (\S\ref{sec:modeling:agnemission:intrinsicblr}). We also consider two \emph{extrinsic} effects (\S\ref{sec:modeling:agnemission:extrinsic}), obscuration by an optically thick torus and dust attenuation along the line of sight.

\subsubsection{\texttt{qsosed}}\label{sec:modeling:agnemission:qsosed}

For our AGN SEDs we have opted not to use the standard \citet{Shakura1973} disc models, as these have well known issues. Firstly, observations show ubiquitous, non-thermal, X-ray emission \citep{Elvis1994, Lusso2016}, which cannot be produced by a standard disc as these are too cool. Secondly, the standard models do not match the detailed spectral shapes seen in the optical/UV \citep{Antonucci1989, Lawrence2018}, showing instead a down-turn in the UV before the expected disc peak \citep{Laor2014, Cai2023}. This down-turn appears to connect across the extreme UV to an upturn in the X-ray, lying above the ubiquitous non-thermal X-ray tail \citep{Laor1997, Porquet2004}, generally dubbed the soft X-ray excess; the origin of which is a matter of some debate \citep{Crummy2006, Petrucci2013, Done2012}. More seriously, the standard \citet{Shakura1973} model predicts variability time-scales intrinsic to the disc on the order several thousands of years, which is in direct conflict with observations \citep[e.g][]{Noda2018, Stern2018, Hernandez2020, Yao2023, Neustadt2024}. Yet, observations do show strong optical/UV emission, and so there must be some optically thick material tapping a significant fraction of the available accretion power. There is likely a disc, but with properties quite different from the standard models.

Hence, we choose to use the modified disc models, which assumes a different vertical structure such that it does not completely thermalise \citep{Rozanska2015, Jiang2020},
implemented with the \qsosed\ code \citep{KD18}; a simplification of the more general \agnsed\ model. These consider a flow radially stratified into three distinct regions: a standard outer disc, an intermediary optically thick non-standard disc-like structure, and then an inner optically thin geometrically thick hot plasma. Here we give a brief overview of the model, but refer the reader to \citet{KD18} for details. For convenience we will also use notation standard to the accretion field. Here $r$ implies dimensionless gravitational radius, related to physical radius $R$ through $R = r R_{G}$ where $R_{G} = GM/c^{2}$. It is also convenient to express the mass-accretion rate through the flow, $\dot{M_{\bullet}}$, scaled by the Eddington mass accretion rate, $\dot{M}_{\rm{Edd}}$, such that $\dot{m} = \dot{M}_{\bullet}/\dot{M}_{\rm{Edd}}$ is dimensionless. This is related to the standard Eddington luminosity through $L_{\rm{Edd}} = \epsilon_{r} \dot{M}_{\rm{Edd}} c^{2}$, where $\epsilon_{r}$ is the spin dependent efficiency at which accretion power is transformed to radiative power.


\qsosed/\agnsed\ start by assuming a standard, relativistic, emissivity profile throughout the entire flow \citep{Novikov1973}. This gives a radial temperature profile, which goes as $T_{NT}^{4}(R) \propto R^{-3} f(R)$, where $f(R)$ describes the radial disc structure in the Kerr metric \citep{Page1974}. The overall accretion power available at each radius (and so overall normalisation of $T_{NT})$ depends purely on the black hole mass, $M_{\bullet}$, mass-accretion rate through the flow, $\dot{M}_{\bullet}$, and efficiency $\epsilon_{r}$. The \qsosed\ and \agnsed\ models use this to set the overall available power, giving energetically self-consistent SED models.

Throughout the flow is subdivided into annular bins of width $\Delta R$ and temperature $T_{NT}(R)$. In the outermost regions of the flow, where the disc is relatively cool, \qsosed/\agnsed\ assume the flow does thermalise into something that looks like a standard \citet{Shakura1973} disc. In this case each radial annulus can be assumed to emit like a black-body, $B_{\nu}(T_{NT}(R))$ with a luminosity given by $2 \times 2\pi R \Delta R \sigma T_{NT}^{4}(R)$, where the extra factor $2$ comes from the disc emitting from both sides and $\sigma$ is the Stefan-Boltzmann constant. Hence, integrating this from the outer disc $r_{\rm{out}}$ to some radius $r$ gives a multi-colour black body SED shape.

As we move further into the flow \qsosed/\agnsed\ assume we reach a region where the disc no longer thermalises, with the transition radius given by $r_{w}$. As the disc temperature increases it will eventually cross the Hydrogen ionisation instability, which gives a dramatic change in disc opacity \citep{Cannizzo1992a}. This could lead to a local instability, giving a vertical disc structure somewhat different from the standard models \citep{Rozanska2015, Jiang2020}, and a potential solution to the variability issue in standard models \citep{Hagen2024a}. In \qsosed/\agnsed\ this is modelled as a warm optically thick plasma overlaying a passive disc, following the work of \citet{Petrucci2013, Petrucci2018}, which can form if the dissipation region moves higher into the photosphere of the disc. Here the seed photons are tied to the underlying disc, such that they have a temperature given by $T_{NT}(R)$ within each annulus. They are further assumed to form a black-body spectrum, which then Compton scatters through the photosphere, modifying the emitted spectrum from each annulus to a Comptonised black-body, calculated using the {\sc nthcomp} code \citep{Zdziarski1996, Zycki1999}. This gives a component in the SED that links across the unobservable EUV; providing a possible explanation for the UV turn-over and soft X-ray excess \citep{Mehdipour2011, Mehdipour2015, Done2012}. Physically this requires the Comptonising photosphere to be optically thick, $\tau >> 1$, with a covering fraction of unity. In order to match to the soft X-ray excess, it also requires an electron temperature, $kT_{e, w}$, of roughly 0.2-0.3\,keV. Within \qsosed, and so for our simulated SEDs, this is simply fixed at $kT_{e, w} = 0.2$\,keV. The corresponding photon index, $\Gamma_{w}$, is also fixed throughout to $2.5$, which is typically seen in the data for bright AGN.

Further in, close to the black hole, the flow no longer forms a disc. Instead, \qsosed\ assumes that below a radius $r_{h}$ it truncates into an optically thin geometrically thick flow \citep{Narayan1994, Liu1999}, extending down to the innermost stable circular orbit, $r_{\rm{isco}}$, and is responsible for forming the X-ray tail. It is assumed that the cooler photons emitted from the optically thick disc structure enter the corona, and Compton up-scatter into the non-thermal X-ray tail. The corona itself is powered by accretion, and for simplifying reasons \qsosed\ assumes a standard \citet{Novikov1973} emissivity to give the total power dissipated within the corona, $L_{\rm{diss}}$. The total observed coronal power also has a contribution from the seed photons entering the corona, $L_{\rm{seed}}$, such that the total power from the X-ray corona is given by $L_{\rm{hot}} = L_{\rm{diss}} + L_{\rm{seed}}$. As with the warm Comptonising disc-like region the spectral shape is calculated using {\sc nthcomp}, and assumed to have a single electron temperature, which in \qsosed\ is fixed at the typical value of $kT_{e, h} = 100$\,keV. The corresponding spectral index, $\Gamma_{h}$, is then calculated off the Compton balance \citep{Beloborodov1999} (i.e. increasing the power dissipated in the corona heats it, giving a harder spectrum, while increasing the seed photon power seen by the corona cools it, giving a softer spectrum). This requires an assumption regarding the geometry, which for simplifying reasons is set to a sphere surrounding the black hole (see \citealt{Hagen2023} Appendix C on calculating the seed photon power).

The total SED model can then be summed up as follows. For $r_{\rm{out}} \geq r > r_{w}$ the flow is assumed to emit like a standard disc, giving a multi-colour black-body. For $r_{w} \geq r > r_{h}$ the disc no longer thermalises, giving instead a warmly Comptonised component to the SED, with $\Gamma_{w}$ and $kT_{e, w}$ fixed at $2.5$ and $0.2$\,keV respectively within \qsosed. Then for $r_{h} \geq r > r_{\rm{isco}}$ the flow forms a hot X-ray plasma, Compton up-scattering seed photons from the disc into the high energy X-ray tail, with $kT_{e, h}=100$\,keV and $\Gamma_{h}$ calculated from the Compton balance.

\begin{figure}[!h]
    \centering
    \includegraphics[width=1\columnwidth]{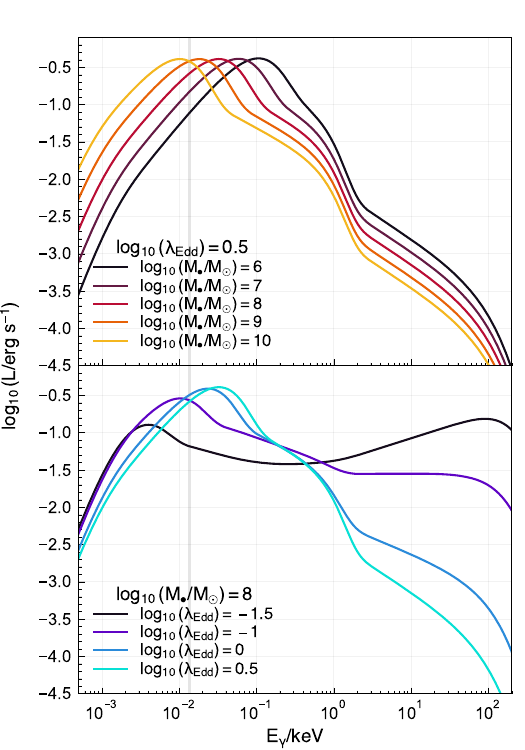}
    \caption{The isotropic bolometric luminosity normalised spectral energy distributions for a small sample of \qsosed\ models. The vertical line delimits energy at which photons are able to ionise Hydrogen. The top-panel shows models with different SMBH mass but assuming $\lambda_{\rm Edd}=1$ while the bottom panel varies $\lambda_{\rm Edd}$ but fixes $M_{\bullet}=10^{8}\ {\rm M_{\odot}}$.}
    \label{fig:modelling:qsosed_spectra}
\end{figure}

A small set of example spectra from the \qsosed\ model, normalised to have $\Lbolagn=1$, are shown in Figure \ref{fig:modelling:qsosed_spectra}. The top-panel of Figure \ref{fig:modelling:qsosed_spectra} shows the impact of changing the SMBH mass at fixed Eddington ratio $\lambda_{\rm Edd}=1$. The key change here is to shift the peak of the emission, which for $\Mbh=10^{9}$ corresponds roughly to the Hydrogen ionisation energy, to lower energies with increasing $M_{\bullet}$. The relative normalisation of the spectrum in the optical also increases by $\approx$ 1.5 dex as $\Mbh=10^{6}\to 10^{9}$; as we will see subsequently this has a significant impact on the ionising photon bolometric correction and H$\alpha$ equivalent width. This behaviour arises from the reduction in the temperature of the accretion disc as the mass increases, shifting the peak of the spectrum to lower energies. Naively one might expect the inverse of this, as the mass sets the potential, however as the potential increases so does the overall surface area of the accretion disc. This increases the net radiation the accretion flow can emit, hence cooling it more efficiently. In the balance between the increased potential and increased surface area, the area wins, leading to an overall reduction in temperature.

The bottom-panel of Figure \ref{fig:modelling:qsosed_spectra} instead shows the impact of varying $\lambda_{\rm Edd}$ keeping the SMBH mass fixed ($\Mbh=10^8$). The impact here is more complex. First, the relative optical luminosity remains roughly constant. However, the peak emission shifts again, this time increasing with $\lambda_{\rm Edd}=1$. However, the shape of the UV and X-ray emission also changes dramatically. This somewhat more complex behaviour is a product of the assumptions going into {\sc qsosed} \citep[see e.g][for a more detailed overview]{Mitchell2023}. \qsosed\ hardwires the total power in the hot Comptonisation component to $0.02\,L_{\rm{Edd}}$, as observations often show X-ray emission limited to $\sim 0.02 - 0.04\,L_{\rm{Edd}}$ over a range of mass and mass-accretion rate \cite[e.g][]{Jin2012a}. Hence, as $\lambda_{\rm Edd}$ decreases, an increasing fraction of the available accretion power is dissipated within the hot corona. In the context of \qsosed\ this manifests itself as the hot corona extending radially outwards. This then leads to a reduction in the overall power emitted in the disc/warm corona, leading to reduced optical/UV emission, as well as the peak shifting to lower energies since the temperature of the inner disc is now also lower. The change in the spectral shape of the X-ray emission comes from the Compton balance, i.e the balance of accretion power heating the free electrons within the plasma and seed photons from the disc cooling the same electrons. As the disc power reduces, there are fewer seed photons, and so the corona does not cool as efficiently, leading to a harder (rising) spectrum.

We note that throughout this paper we only calculate SEDs with $\lambda_{\rm Edd}>0.03$, as below this the flow is expected to fully transition to an inefficient X-ray plasma (i.e no optically thick disc structure), as seen in both monitoring of changing-look AGN \citep{Noda2018, Panda2024} and in the wider AGN population through single epoch SEDs \citep{Hagen2024b}. We also note that above $\lambda_{\rm Edd} \sim 1$ we expect the inner part of the accretion disc to deviate from the thin disc approximation, puffing up into a `slim' disc due to the increased radiation pressure. Advection/wind losses should then limit the surface luminosity to the local Eddington limit, effectively altering the emissivity profile of the disc \citep{Abramowicz1988, Kubota2019}. 


\begin{figure}
    \centering
    \includegraphics[width=1\columnwidth]{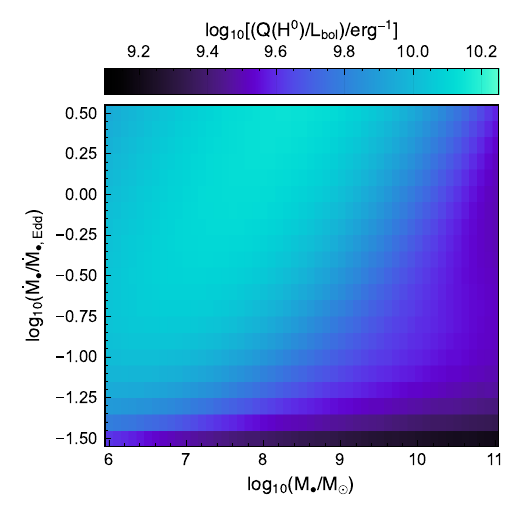}
    \includegraphics[width=1\columnwidth]{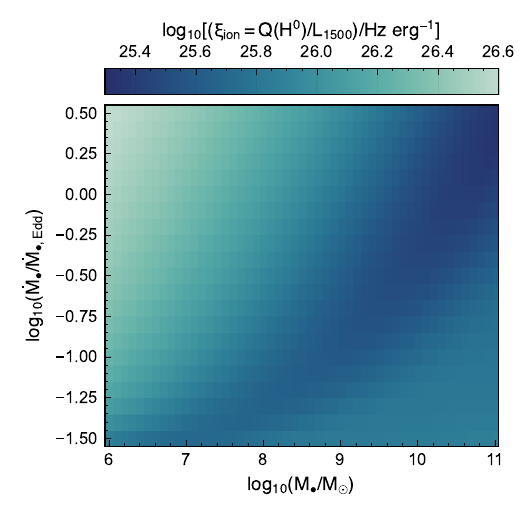}
    \caption{The H\textsc{i} ionising photon bolometric correction (top, $Q(H^0)/\Lbol$) and the \emph{intrinsic} ionising photon production efficiency (bottom, $\xi_{\rm ion}=Q(H^0)/L_{1500}$) predicted by \qsosed\ as a function of SMBH mass and accretion rate.}
    \label{fig:modelling:ionising_bolometric_correction}
\end{figure}

Since, in this work, we are interested in the H$\alpha$ emission from SMBHs it is useful to also explore predictions for the Hydrogen ionising photon luminosities ($Q(H^0)$) as a function of both SMBH mass and accretion rate or Eddington luminosity. The top panel of Figure \ref{fig:modelling:ionising_bolometric_correction} shows the Hydrogen ionising photon bolometric correction (i.e. $Q(H^0)/L_{\rm bol}$) as a function of SMBH mass and accretion rate for a large grid of models. This mirrors the trends seen in Figure \ref{fig:modelling:qsosed_spectra} revealing variations of up to $\approx 0.8$ dex across the range of SMBH masses and accretion rates. For example, at fixed $\lambda_{\rm Edd}=1$, the relative ionising photon luminosities reaches a peak at $\log_{10}(M_{\bullet}/{\rm M_{\odot}})\approx 8.5$. At lower $\lambda_{\rm Edd}$ this peak shifts to lower masses. Figure \ref{fig:modelling:ionising_bolometric_correction} also shows the \emph{intrinsic} ionising photon production efficiency $\xi_{\rm ion}=Q(H^0)/L_{1500}$ similarly highlighting that the \qsosed\ model predicts a wide range of values depending on the mass and accretion rate.

\subsubsection{The Intrinsic BLR Emission}\label{sec:modeling:agnemission:intrinsicblr}

Since our goal is to model the Hydrogen recombination line emission from the broad-line emitting regions we next need to consider what fraction of the ionising emission from the disc is reprocessed and how ionising photons are reprocessed into Hydrogen recombination line photons. 

First, the broad-line emitting region is not expected to entirely cover the central disc, i.e. some ionising radiation is able to directly escape; either entirely, or to illuminate clouds at larger radii, with lower density, producing narrow-line emission. We parameterise this as the BLR covering fraction $f_{\rm cov}$ and explore how changes to its value $\in[0.1, 1.0]$ affect our predictions. 

The next challenge is modelling the conversion of ionising to Hydrogen recombination line photons, specifically H$\alpha$. At the low-densities ($n_{H}< 10^{4}\ {\rm cm^{-3}}$) of typical H\textsc{ii} regions Case B recombination, assuming $T=10^{4}\ {\rm K}$, predicts:
\begin{equation}
\frac{L_{\rm H\beta}}{\int_{\nu_0}^{\infty}L_{\nu}\ (h\nu)^{-1}\ {\rm d}\nu} = h\nu_{\rm H\beta}\frac{\alpha_{\rm H\beta}}{\alpha_{B}} \approx 4.8\times 10^{-13}\ {\rm erg}
\end{equation}
where $L_{\rm H\beta}$ is the H$\beta$ luminosity, $\alpha_{B}$ is the Case B recombination coefficient. This can be re-expressed as a conversion for H$\alpha$ by simply using the expected Case B relative line intensities $L_{H\alpha}/L_{H\beta} \approx 2.86$ (again for $T=10^{4}\ {\rm K}$). 

However, BLRs are characterised by high-densities $n_{H}\sim 10^{10}\ {\rm cm^{-3}}$ where collisional processes become increasingly important resulting in deviations from the low-density Case B prediction. While a more accurate conversion can be obtained with full photoionisation modelling this requires additional assumptions, such as the exact ionisation parameter, hydrogen density, and column density. Since, in this work, we are more interested in obtaining rapid predictions, we adopt the results of the homogeneous BLR models presented in \citet{Osterbrock}. For temperatures of 10,000 K (14,000 K), corresponding to ionisation parameters of $10^{-4}$ ($10^{-2}$), these models predict $2.2\times$ ($0.43\times$) as many H$\alpha$ photons are produced per ionising photon relative to the low-density Case B recombination predictions. For the purpose of this work we assume the larger of these values, but caution that this is a limiting factor of this analysis. In reality this conversion is, even in simple constant density models, sensitive to the shape and intensity (i.e. ionisation parameter) of the incident radiation, the hydrogen density, the chemical composition, and the column density of gas, and can reach values $>10\times$ the low-density Case B prediction, or entirely suppress H$\alpha$ emission (for example at large column densities). This is important as there is now growing evidence \citep[e.g.][]{Inayoshi2025, Ji2025, DEugenio2025, Naidu2025, Rusakov2025} that at least some early AGN are surrounded by large column densities and covering fractions of gas which is highly turbulent.

\begin{figure}
    \centering
    \includegraphics[width=1\columnwidth]{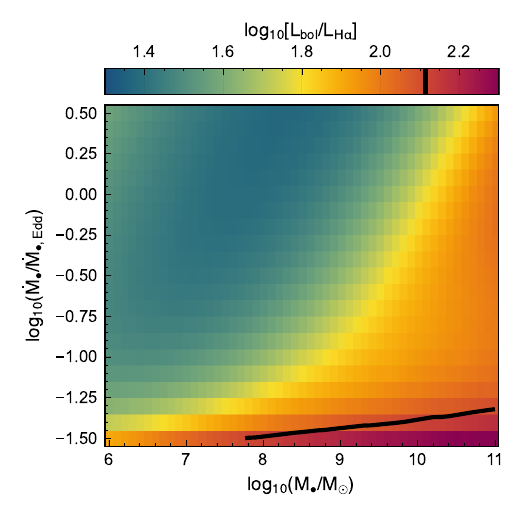}
    \includegraphics[width=1\columnwidth]{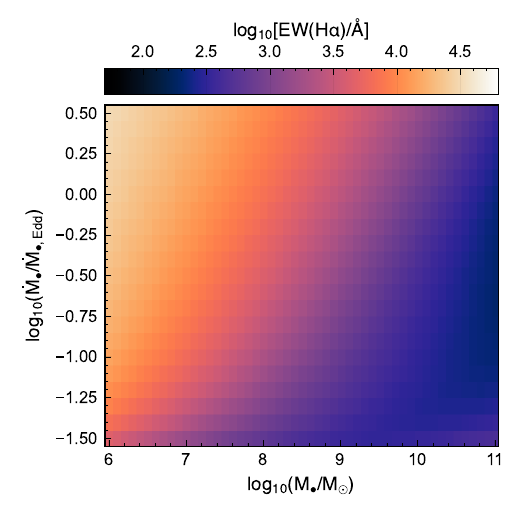}
    \caption{The H$\alpha$ bolometric correction (top, plotted as $L_{\rm bol}/L_{\rm H\alpha}$ and equivalent width (bottom) predicted by \qsosed\ as a function of SMBH mass and accretion rate assuming a constant conversion of ionising photons to H$\alpha$ photons, based on simple photoionisation modelling, and a covering fraction $\fcov=1$. The black line and contour on the bolometric correction panel denote the widely used \citet{Stern2012} correction.}
    \label{fig:modelling:qsosed_halpha}
\end{figure}

Assuming this conversion we can predict the H$\alpha$ bolometric correction as a function of the SMBH mass and accretion rate. This is shown in the top-panel of Figure \ref{fig:modelling:qsosed_halpha}. Since we have assumed a constant conversion this is simply a scaling of the ionising photon bolometric correction. We also mark a common utilised \citet{Stern2012} bolometric correction ($L_{\rm bol}/L_{\rm H\alpha}=130$), which utilised the \citet{Richards2006} AGN SED, and is widely utilised in the literature. This reinforces the \qsosed\ prediction that the bolometric correction should vary with the SMBH mass and accretion rate (and thus luminosity) and thus the assumption of a single correction is inappropriate and will introduce an additional source of uncertainty and bias. While SMBHs accreting around the Eddington limit at $\Mbh<10^9$ have predicted bolometric corrections smaller than the widely adopted value, here we assume a covering fraction of unity; for smaller covering fractions more BHs in the range $\Mbh=10^7-10^9$ will have bolometric corrections consistent with this value.

The bottom panel of Figure \ref{fig:modelling:qsosed_halpha} shows the predicted H$\alpha$ equivalent width (EW), again assuming a covering fraction $f_{\rm cov}=1$. Here the continuum level used to calculate the EW is based solely on the disc SED and so does not include the contribution of nebular continuum emission which would require full photoionisation modelling and is beyond the scope of this article, nor does it include the contribution of stars. Thus the EW is simply $\propto Q(H^{0})/L_{\lambda}(6536{\rm\AA})$. Since a reduction in the ionising photon luminosity is to some extent balanced by an increase in the optical luminosity the EW shows a stronger dependence on the mass and accretion rate than the bolometric correction. For SMBHs accreting at the Eddington limit this suggests the EW should decrease from $\sim 5000{\rm\AA}$ for BHs with $\Mbh=10^7$ to $<1000{\rm\AA}$ for BHs with $\Mbh>10^9$.

\subsubsection{Extrinsic effects}\label{sec:modeling:agnemission:extrinsic}

We also consider two extrinsic effects which affect the observability of the BLR and the H$\alpha$ emission.

First, the disc and BLR of active SMBHs are not always observable due to the presence of an obscuring torus of material along the line-of-sight. Since we are focussed only on broad-line AGN we account for this effect with an additional parameter, $f_{\rm BLR}$, the fraction of AGN where observable broad-line emission emerges. Here we explore $f_{\rm BLR}\in [0.1, 1.0]$. In practice we simply ignore the contribution $f_{\rm BLR}$ SMBHs when considering distribution functions, such as the H$\alpha$ luminosity function. Since this is sensitive to the sampling we run 1000 realisations and take the median result.

Secondly, AGN are embedded in galaxies and their emission might naturally be attenuated by dust along the line-of-sight. FLARES utilises a complex dust-geometry such that individual star (and potentially SMBH) particles have independent dust optical depths based on the line-of-sight surface density of metals \citep[see:][]{FLARES-II, FLARES-XII}, assuming a constant dust-to-metal ratio. While this yields good results for starlight we would expect it to fail in the vicinity of AGN where we might naturally expect a lower dust-to-metal ratio due to increased destruction of AGN. Indeed, if the FLARES model is naively applied, the resulting extinction of the AGN is enough to dramatically diminish their contribution to the emission from virtually all galaxies. For this reason, in this work we do not apply the existing FLARES dust modelling for AGN emission. In the context of the H$\alpha$ luminosity function presented in \S\ref{sec:results:ha:half} we explore AGN dust optical depths parameterised as a scaling of the integrated stellar optical depth $\tau_{\bullet}/\tau_{\star} \in\{0., 0.5, 1, 2\}$. 

\subsection{Stellar Emission}\label{sec:modeling:emission}

While the main focus of this work is the emission from AGN we also show comparisons with stellar H$\alpha$ emission. The emission from stars in FLARES (and their associated H\textsc{ii}) are described in \citet{FLARES-II}. In short, every star particle in each galaxy is associated with a pure stellar spectra based on its age, metallicity, and mass, assuming version \texttt{2.2.1} of the Binary Population and Spectral \citep[BPASS][]{BPASS2.2.1} stellar population synthesis model and a \citet{ChabrierIMF} initial mass function. Each star particle is also associated with a H\textsc{ii} region modelled using the \texttt{cloudy} \citep{Cloudy17.02} photoionisation code assuming the same metallicity as the star particle. Here we assume an ionisation parameter of $U=0.01$ (referenced at $1 {\rm Myr}$ and $Z=0.01$), hydrogen density $n_{H}=100\ {\rm cm^{-3}}$, and include Orion-like grains which scale with the metallicity. As noted above, dust attenuation is modelled on a particle-by-particle basis by converting the line-of-sight density of metals to an optical depth.

%% file: sections/results.tex
\section{Results}\label{sec:results}

Using the methodology described in \S\ref{sec:modeling} we have calculated the ionising photon luminosities and H$\alpha$ emission for all SMBHs in FLARES. Here we present our predictions for both.  

\subsection{Ionising photon emissivities}\label{sec:results:ionising}

\subsubsection{Bolometric correction}

\begin{figure}
    \centering
    \includegraphics[width=1\columnwidth]{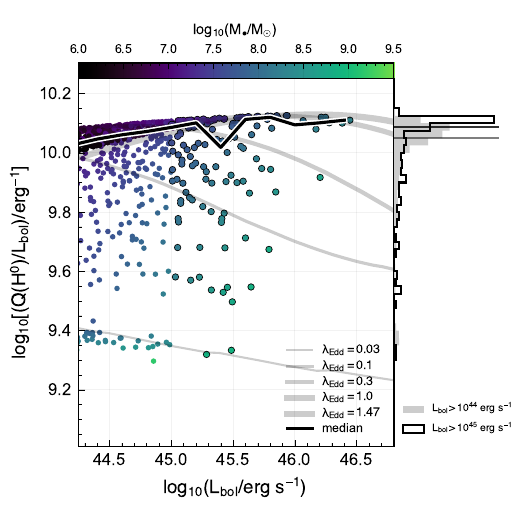}
    \caption{The relationship between the bolometric luminosity and the ratio of ionising photon luminosity to the bolometric luminosity ($Q(H^{0})/\Lbol$) predicted by FLARES at $z=5$. The dark outlined line shows the binned median. The faint lines show predictions from the \qsosed\ model for different Eddington ratios as a function of SMBH mass. The right-hand panel shows a histogram of the $\lambda_{\rm Edd}$ for $\Lbol>10^{44}$ (faint, filled histogram) and $\Lbol>10^{45}$ (dark, un-filled) with the faint and dark horizontal lines denoting the $\Lbol>10^{44}$ ($>10^{45}$) median values respectively.}
    \label{fig:modelling:bolometric_luminosity-ionising_photon_luminosity}
\end{figure}

We first explore the relationship between the bolometric luminosity and the Hydrogen ionising photon luminosity $Q(H^{0})$. Figure \ref{fig:modelling:bolometric_luminosity-ionising_photon_luminosity} shows this, expressed in terms of the ratio of ionising photon luminosity to the bolometric luminosity ($Q(H^{0})/\Lbol$), or the bolometric correction. Here the solid curved lines denote the bolometric corrections for SMBHs accreting at various Eddington ratios, including the maximum permitted in the FLARES/EAGLE physics model ($\lambda_{\rm Edd}=h^{-1}\approx 1.48$). The predicted FLARES corrections range over $\log_{10}[(Q(H^{0})/\Lbol)/{\rm erg}^{-1}]\approx [9.3, 10.1]$; this demonstrates that AGN should not all have the same bolometric correction. However the median value is $\approx 10^{10.1}$, i.e. close to the maximum, for both our liberal and conservative samples. The median ratio increases slightly ($\approx 0.1$ dex) from $\Lbol=10^{44}\to 10^{46}$.

\subsubsection{Ionising photon production efficiency}

\begin{figure}
    \centering
    \includegraphics[width=1\columnwidth]{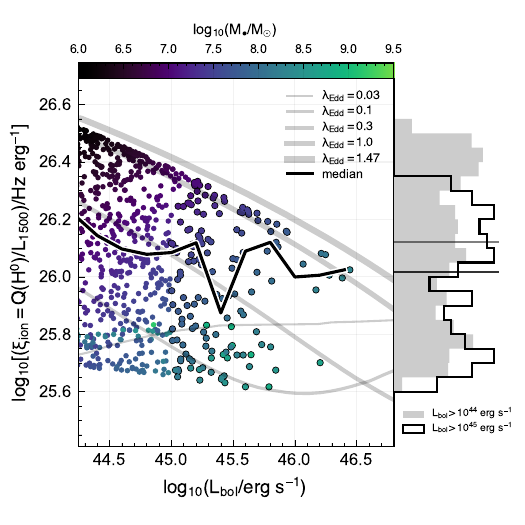}
    \caption{The relationship between the bolometric luminosity and the ionising photon production efficiency ($\xi_{\rm ion}=Q(H^{0})/L_{1500}$) predicted by FLARES at $z=5$. The dark outlined line shows the binned median. The faint lines show predictions from the \qsosed\ model for different Eddington ratios as a function of SMBH mass. The right-hand panel shows a histogram of the $\lambda_{\rm Edd}$ for $\Lbol>10^{44}$ (faint, filled histogram) and $\Lbol>10^{45}$ (dark, un-filled) with the faint and dark horizontal lines denoting the $\Lbol>10^{44}$ ($>10^{45}$) median values respectively.}
    \label{fig:modelling:bolometric_luminosity-production_efficiency}
\end{figure}

Next in Figure \ref{fig:modelling:bolometric_luminosity-production_efficiency} we show our predictions for the ionising photon production efficiency $\xi_{\rm ion}$, alongside tracks at fixed Eddington ratio. Compared to the bolometric corrections, production efficiencies span a wider range $\log_{10}(\xi_{\rm ion}/{\rm Hz\ erg^{-1}})=25.4-26.5$ and have a larger scatter ($\sigma[\log_{10}(\xi_{\rm ion}/{\rm Hz\ erg^{-1}}]=0.23$). Across the bolometric luminosities we study the average $\xi_{\rm ion}$ is relatively flat with a value $\xi_{\rm ion}/{\rm Hz\ erg^{-1}}\approx 10^{26}$. This is $\approx 0.6$ dex larger than that predicted due to stars alone \citep{FLARES-XIII} and compatible with the largest values seen in observations.  

\subsubsection{Relative contribution}\label{sec:results:ionising:rel}

\begin{figure}
    \centering
    \includegraphics[width=1\columnwidth]{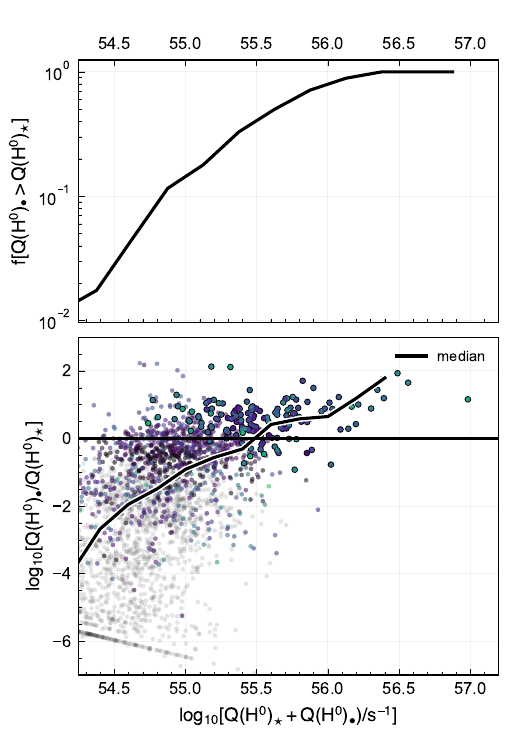}
    \caption{The bottom-panel shows the relationship between the total ionising luminosity and the ratio of stellar and AGN ionising luminosities. Colour-coded points meet our liberal selection criteria ($\Lbol>10^{44}$) while points that are also outlined meet our conservative criteria ($\Lbol>10^{45}$) as well. Grey points have SMBH bolometric luminosities $<10^{44}\ {\rm erg\ s^{-1}}$. The thin horizontal line denotes the boundary between AGN and stellar dominated and the outlined line denote the binned median. The top-panel instead shows the fraction of galaxies where the SMBH dominates the production of ionising photons.}
    \label{fig:modelling:stellar_agn_ionising_relative}
\end{figure}

Next, we explore the relative contributions of AGN and stars to the intrinsic ionising photon luminosities of galaxies. The bottom-panel of Figure \ref{fig:modelling:stellar_agn_ionising_relative} shows the relative contribution of AGN and stars ($Q(H^0)_{\bullet}/Q(H^0)_{\star}$) as a function of the \emph{total} ionising photon luminosity. The median contribution of SMBHs to the ionising photon luminosity increases rapidly with total luminosity. For galaxies with $Q(H^0)\sim 10^{54.5}\ {\rm s^{-1}}$ SMBHs, on average, only contribute around $1\%$ of the total ionising luminosities of individual galaxies. In contrast, above $Q(H^0)=10^{55.5}\ {\rm s^{-1}}$, SMBHs, on average, dominate the production of ionising photons (i.e. $Q(H^0)_{\bullet}/Q(H^0)_{\star}>1$). It is also important to note that at fixed ionising luminosity there is significant scatter in the relative contribution of SMBHs and stars. Even at the faintest luminosities considered there are objects dominated by their SMBH. This can be seen more clearly in the top-panel of Figure \ref{fig:modelling:stellar_agn_ionising_relative}; this shows the fraction of galaxies where the AGN dominates the ionising photon output. This rises from $\approx 0.03$ at $Q(H^0)=10^{54.5}\ {\rm s^{-1}}$ to unity at $Q(H^0)=10^{56.5}\ {\rm s^{-1}}$, i.e. in all objects with $Q(H^0)>10^{56.5}\ {\rm s^{-1}}$ the ionising emission is dominated by the SMBH.

\subsubsection{Luminosity function}\label{sec:results:ionising:lf}

\begin{figure}
    \centering
    \includegraphics[width=1\columnwidth]{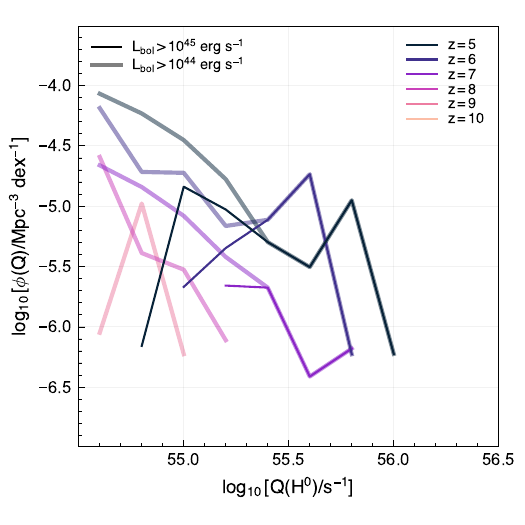}
    \caption{The ionising photon luminosity function at $z=[5, 10]$ predicted by FLARES. Faint thick lines denote our liberal ($\Lbol>10^{44}$) while the thin line denotes our conservative sample ($\Lbol>10^{45}$). }
    \label{fig:modelling:ionising_luminosity_function}
\end{figure}

We next, in Figure \ref{fig:modelling:ionising_luminosity_function}, show the predicted ionising photon luminosity function at $z=5\to 10$, showing results for both our conservative ($\Lbol>10^{45}$) and liberal selections ($\Lbol>10^{44}$). These luminosity functions largely mimic the trends seen in the bolometric luminosity function, except the bright-end evolves slightly more strongly since more (bolometrically) luminous SMBHs tend to have larger ionising-to-bolometric ratios. 

\subsection{Hydrogen recombination lines}\label{sec:results:ha}

We now focus our attention on predictions for the H$\alpha$ properties of galaxies including the relative contribution of stars and AGN, the luminosity function and equivalent width (EW) distribution.

\subsubsection{Hydrogen-$\alpha$ luminosity function}\label{sec:results:ha:half}

We begin by exploring predictions for the BLAGN H$\alpha$ luminosity function (H$\alpha$LF), focussing, initially, on $z=5$ where we have the best statistics and observational comparison samples. 

\begin{figure*}
    \centering
    \includegraphics[width=1\columnwidth]{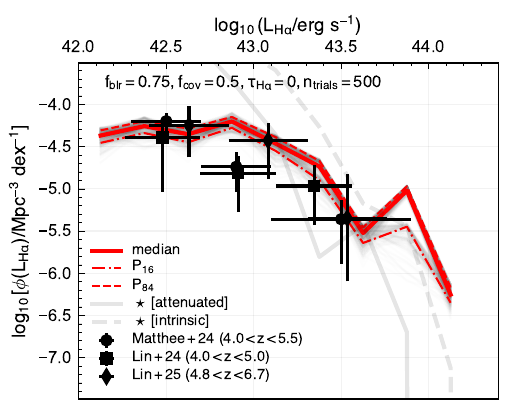}
    \includegraphics[width=1\columnwidth]{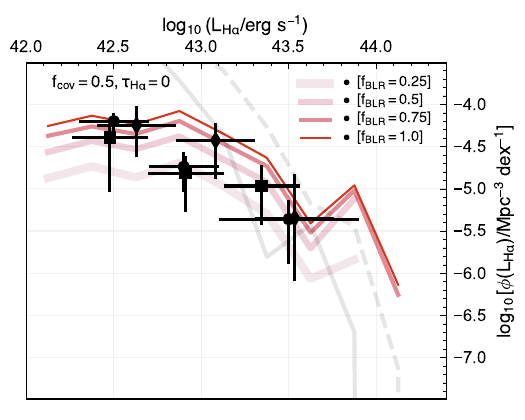}
    \includegraphics[width=1\columnwidth]{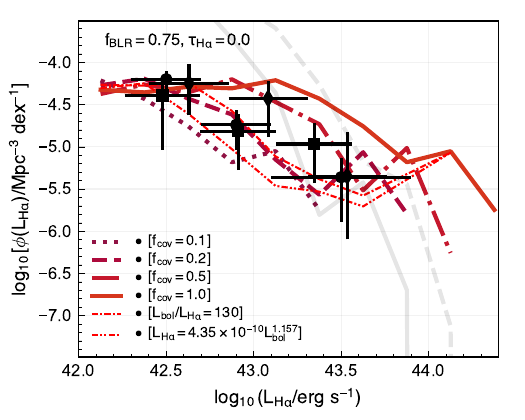}
    \includegraphics[width=1\columnwidth]{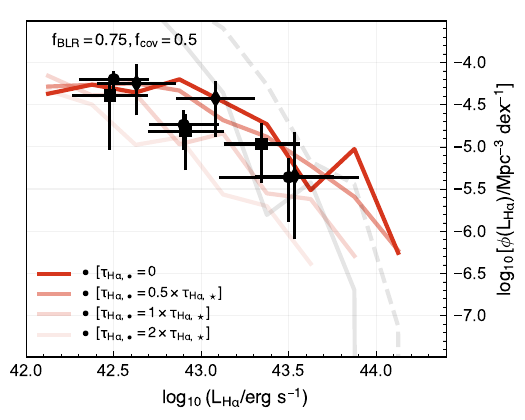}
    \caption{The predicted AGN H$\alpha$ luminosity function (LF) at $z=5$, showing the impact of different modelling choices. Since $f_{\rm BLR}<1$ requires sampling the lines show the median LF based on 500 realisations. The top-right panel shows 500 individual realisations for one value of $f_{\rm BLR}$ ($=0.75$) alongside the median and 16$^{\rm th}$ and 84$^{\rm th}$ percentiles. The top-right-panel shows predictions assuming different values of $f_{\rm BLR}$ (assuming $f_{\rm cov}=0.5$). The bottom-left-panel shows predictions assuming several different values of the covering fraction ($f_{\rm cov}\in\{0.1, 0.2, 0.5, 1.0\}$) (assuming $f_{\rm BLR}=0.75$). The bottom-right-panel shows predictions for different dust optical-depths relative to the stellar optical depth. The thick grey faint solid and dashed lines in each panel show the intrinsic and dust attenuated stellar H$\alpha$ luminosity functions. The points show the observational constraints from \citet{Matthee23} and \citet{Lin24}.}
    \label{fig:results:halpha_luminosity_function}
\end{figure*}

Here we focus our observational comparisons on the studies of \citet{Matthee23} and \citet{Lin24} who have both published broad-line H$\alpha$ emitter (BHAE) luminosity functions. \citet{Matthee23} identified 20 BHAEs at $4<z<6$ using deep \emph{JWST}/NIRCam imaging and wide field slitless spectroscopy (WFSS) from the EIGER \citep{EIGER} and FRESCO surveys \citep{FRESCO}. Similarly, \citet{Lin24} identified 16 BHAEs at $4<z<5$ using NIRCam WFSS obtained over wide area ($\sim 275\ {\rm arcmin}^2$) by the ASPIRE programme, obtaining constraints of the H$\alpha$ LF consistent with \citet{Matthee23}. More recently, \citep{Lin2025} identified 13 BLAGN from the COSMOS-3D survey. These observational constraints are included on each panel of Figure \ref{fig:results:halpha_luminosity_function} making the assumption that all the BHAEs are BLAGN. In this Figure we explore the impact of the various modelling assumptions on the predicted H$\alpha$ luminosity function. Each panel on Figure \ref{fig:results:halpha_luminosity_function} also shows both the intrinsic and attenuated H$\alpha$LF for just the stellar component of galaxies. As expected from Figure \ref{fig:results:stellar_agn_relative}, \flares\ is probing the regime where AGN \emph{potentially}, for certain assumption of $\fblr$, $\fcov$ and dust attenuation, dominate the H$\alpha$LF. However, at $\LHa<10^{43}$ the contribution of AGN becomes negligible. 

In the top-panels of Figure \ref{fig:results:halpha_luminosity_function} we explore the impact of changing $\fblr$. For values of $\fblr$ less than unity we simply discard $1-\fblr$ of the AGN; as a consequence any single realisation will be subject to statistical uncertainty since the number of bright AGN is relatively small. To account for this, when modelling $\fblr<1$ we take median value from an ensemble of 500 realisations. To demonstrate the statistical uncertainty resulting from this sampling, in the top-left panel of Figure \ref{fig:results:halpha_luminosity_function} we show the luminosity function for the 500 individual realisations (faint grey lines) alongside the median, and 16$^{\rm th}$ and 84$^{\rm th}$ percentiles. This highlights the additional statistical uncertainty which will increase further for lower $\fblr$ and at higher-redshift, where the total number of sources is smaller. It is important to stress that this is not the only source of statistical uncertainty and that the prediction variance is likely to be significantly larger, likely comparable to the current observational uncertainties. 

In the top-right panel of Figure \ref{fig:results:halpha_luminosity_function} we next show the predicted BLAGN H$\alpha$ luminosity function for different values of $f_{\rm BLR}$ assuming a covering fraction of $0.5$ ($\fcov=0.5$). When we assume all AGN have observable broad-line regions (i.e. $f_{\rm BLR}=1$) \flares\ produces a good match to the faintest observational bins but tends to predict more bright sources than observed. Reducing $f_{\rm BLR}$ simply shifts the predicted H$\alpha$LF to lower densities, resulting in better agreement with observations at the bright end but worse agreement at the faint-end. 

We next explore the impact of changing the covering fraction $f_{\rm cov}$; the bottom-left panel of Figure \ref{fig:results:halpha_luminosity_function} shows the predicted H$\alpha$LF for different choices for the covering fraction ($f_{\rm cov}\in\{0.1, 0.2, 0.5, 1.0\}$), assuming $f_{\rm BLR}=0.75$. Since the impact of $\fcov$ is to simply scale luminosities the impact of reducing $f_{\rm cov}$ is to simply \emph{shift} the luminosity function to lower luminosities. The flatness of the predicted LF at $\LHa<10^{43.5}$ (assuming $\fcov=1$) results in the faint-end normalisation of the LF remaining approximately constant at different $\fcov$. As a consequence a high $\fblr$ ($>0.75$) and low ($\sim 0.2$) $\fcov$ can yield a good fit to both the bright and faint ends of the observed LF. Changing the conversion of the ionising photon to H$\alpha$ luminosities will mimic the impact of $\fcov$. That is, if the conversion is larger than assumed then a smaller covering fraction would be required to produce a good match and vice versa. This panel also shows the result of assuming a constant bolometric correction of $L_{\rm bol}/L_{\rm H\alpha}=130$ \citep{Stern2012} and the luminosity dependent correction of \citet{Richards2006}, assumed by \citet{Kokorev2024} and \citet{Matthee23} respectively to infer the bolometric luminosity function. Both corrections produce a result similar to assuming $\fcov=0.2$, yielding a good agreement with the $z=5$ observations. It is important to note that in our model, since the BLR is assumed to be ionisation bounded a covering fraction of unity implies no ionising photons escape and thus there would be no narrow line emitting region, thus no Type II AGN. However, observations of Type II AGN at high-redshift exist \citet[e.g.][]{Scholtz2023} suggesting that $\fcov<1$, or that the BLR is density bounded.

Finally, in the bottom panel of Figure \ref{fig:results:halpha_luminosity_function}, we explore the impact of dust attenuation. Here we explore the assumption that the dust optical depth for SMBHs is simply proportional to the effective (i.e. galaxy wide) stellar optical depth, assuming $\fblr=0.75$ and $\fcov=0.5$. Like reducing the covering fraction, increasing the optical depth has a stronger effect on the bright end of the luminosity function. In this case, this is because of both the flatness of the faint-end slope but also the tendency for brighter sources to have larger optical depths. Assuming an AGN optical depth the same as the effective stellar optical depth (assuming $\fblr=0.75$ and $\fcov=0.5$) produces a good match to the observations, similar to assuming $\fcov=0.2$ with no dust attenuation. 

\begin{figure*}
    \centering
    \includegraphics[width=2\columnwidth]{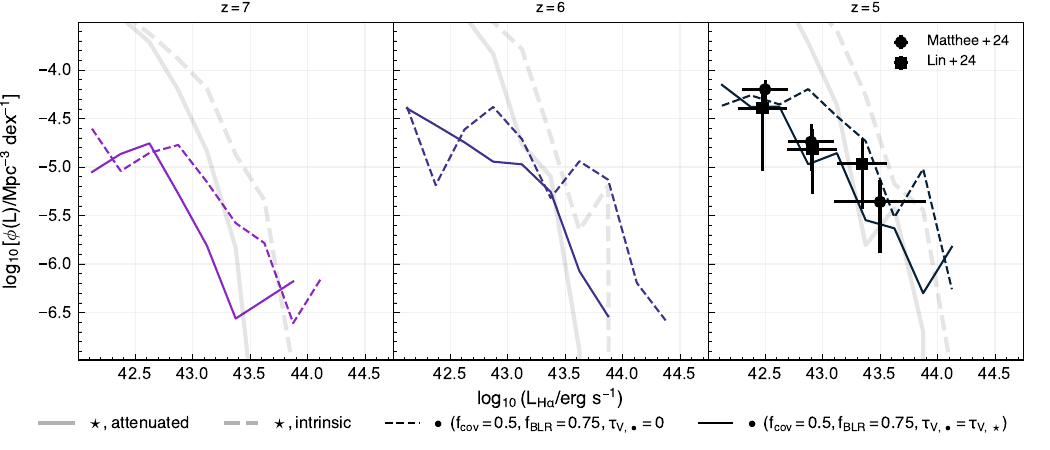}
    \caption{The evolution of the stellar and AGN (assuming $\fblr=0.75$, $f_{\rm cov}=0.5$, $\tau_{\bullet}\in\{0, \tau_{\star}\}$) H$\alpha$ luminosity functions at $z\in\{7,6,5\}$. Also shown at $z=5$ are observations from \citet{Matthee23} and \citet{Lin24}.}
    \label{fig:results:halpha_luminosity_function_evolution}
\end{figure*}

For completeness, we also show predictions at higher-redshift $5 \le z\le 7$ in Figure \ref{fig:results:halpha_luminosity_function_evolution} for our fiducial set of parameters $\fcov=0.5$, $\fblr=0.75$, and $\tau_{\bullet}=\tau_{\star}$. This reinforces that there exist a set of parameters that produce good agreement between the FLARES predictions and current observational constraints. In terms of the predicted evolution, we see a clear drop $z=5\to 7$ of $\approx 0.5$ dex across the modelled luminosity range.

\subsubsection{Relative contribution of stars and AGN}\label{sec:results:ha:rel}

\begin{figure}
    \centering
    \includegraphics[width=1\columnwidth]{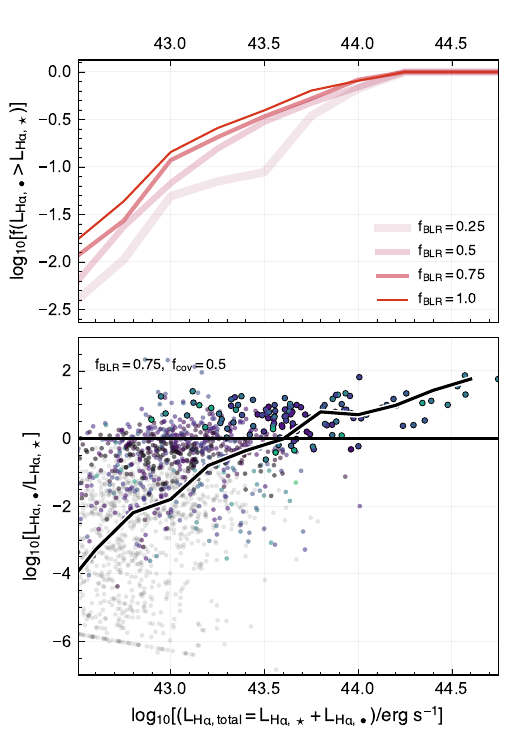}
    \caption{The relative contribution of stars and AGN to the H$\alpha$ emission at $z=5$. The bottom panel relative contribution of accreting SMBHs to the total H$\alpha$ luminosity of galaxies predicted by FLARES assuming a covering fraction of $\fcov=0.5$. Coloured points denote those SMBHs meeting our conservative selection criteria ($M_{\bullet}/{\rm M_{\odot}}>10^{7}$, $L_{\rm \bullet, bol}/{\rm erg\ s^{-1}}>10^{45}$). The top panel shows the fraction of galaxies in each luminosity bin in which the luminosity is dominated by the accreting SMBH.}
    \label{fig:results:stellar_agn_relative}
\end{figure}

Next, we explore the relative contribution of AGN and stars to the H$\alpha$ luminosity. The bottom panel of Figure \ref{fig:results:stellar_agn_relative} is similar to that in Figure \ref{fig:modelling:stellar_agn_ionising_relative} but also includes an assumed BLR fraction ($\fblr$) and covering fraction ($\fcov$), specifically assuming our fiducial set of parameters $\fcov=0.5$, $\fblr=0.75$. While this represents a single realisation it is worth noting that the conclusions do not vary significantly. Like in Figure \ref{fig:modelling:stellar_agn_ionising_relative} the relative contribution increases rapidly with luminosity; above $L_{H\alpha}=10^{43.7}\ {\rm erg\ s^{-1}}$ the H$\alpha$ luminosity is dominated, in the majority of galaxies, by the emission from the AGN. The top-panel of Figure \ref{fig:results:stellar_agn_relative} instead shows the fraction of sources where the AGN dominates the luminosity. Unlike the corresponding figure for the ionising luminosity here we explore alternative values of $\fblr$, using 500 realisations. Reducing $\fblr$ inevitably leads to a smaller fraction of galaxies, at fixed luminosity, being dominated by their SMBH. However, even for $\fblr=0.25$ the H$\alpha$ emission of the brightest galaxies $\LHa>10^{44}$ remain predominantly powered by their AGN.

\subsubsection{Hydrogen-$\alpha$ equivalent width distribution}\label{sec:results:ha:ew}

In \S\ref{sec:results:ha:half} we showed that \flares\ predictions can be brought into agreement with observations for particular choices of the BLR fraction ($\fblr$), covering fraction ($\fcov$), and dust optical depth ($\tauagn$). Specifically the faint-end observations prefer high-values $\fblr$ ($>0.75$) while good overall agreement can be obtained by assuming either $\fcov=0.2$ and no dust or $\fcov=0.5$ and a dust optical depth similar to the effective stellar optical depth. 

However, while the covering fraction and dust optical depth both affect line luminosities in similar ways they will have different impacts on line ratios (e.g. the Balmer decrement: H$\alpha$/H$\beta$) and equivalent widths. 

Dust attenuation preferentially diminishes the shorter-wavelength H$\beta$ emission relative to H$\alpha$, leading to elevated values of the H$\alpha$/H$\beta$ ratio. At present, measurements of the Balmer decrement in high-redshift broad-line AGN remain scarce \citep[though see e.g.][]{Brooks2024}. Furthermore, in contrast to the low-density regime where the intrinsic line ratio is relatively well-defined, the ratio in broad-line regions can vary significantly (typically in the range $\sim$2–10), depending on factors such as gas density, column density, and the ionisation parameter (Vijayan et al., in preparation). These dependencies introduce substantial complexity in the interpretation of Balmer decrement measurements in BLAGN.

The impact of dust and covering fraction on equivalent widths is more complex. When considering only the AGN the impact of reducing the covering fraction will be to reduce the EW, since the line emission reduces but the disc emission stay constant. In reality the decrease in EW is not quite linear, since there is a contribution to the continuum from the nebular continuum emission (which we do not model here), which would also scale with the covering fraction. On the other hand dust should, assuming it is extrinsic to both the disc and BLR, leave the EW unchanged. Like the Balmer decrement, there are relatively few observational constraints of BL AGN H$\alpha$ equivalent widths since often the continuum is not detected sufficiently to enable a robust measurement. However, \citet{Matthee23} note EWs of $\sim 500{\rm\AA}$ in their sample while \citet{Lin24} find EWs $100-2000\ {\rm\AA}$ across their sample. Interpreting EWs is further complicated by the contribution of stellar emission to the continuum which, in most of our sample, is comparable to or larger than the contribution from the AGN itself. 

\begin{figure}
    \centering
    \includegraphics[width=1\columnwidth]{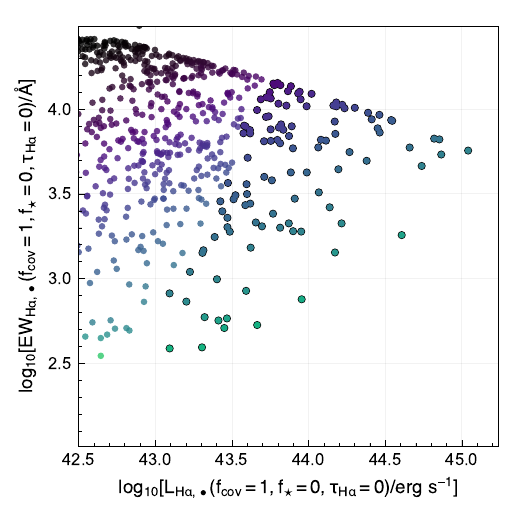}
    \caption{The relationship between the predicted H$\alpha$ luminosity and equivalent width (EW) assuming a covering fraction $\fcov$ of unity, no stellar continuum emission ($f_{\star}=0$), and no dust attenuation ($\tau_{\bullet}=0$).}
    \label{fig:results:agn_ew}
\end{figure}

With these caveats in mind we conduct a brief exploration. Figure \ref{fig:results:agn_ew} shows the relationship between the luminosity and equivalent width for \emph{pure} AGN (i.e. excluding the contribution from stars and nebular continuum emission) assuming a covering fraction of unity. At $\LHa>10^{43}$ we find a broad range of predicted EW values: $EW/{\rm\AA}\sim [300, 2\times 10^{4}]$, driven predominantly by the different SMBH masses, but also accretion rates, as discussed in \S\ref{sec:modeling:agnemission:qsosed}. It is worth noting that the maximum EW tends to decrease with increasing H$\alpha$ luminosity. This simply reflects the fact that for a fixed Eddington ratio the expected EW decreases with SMBH mass.

\begin{figure}
    \centering
    \includegraphics[width=1\columnwidth]{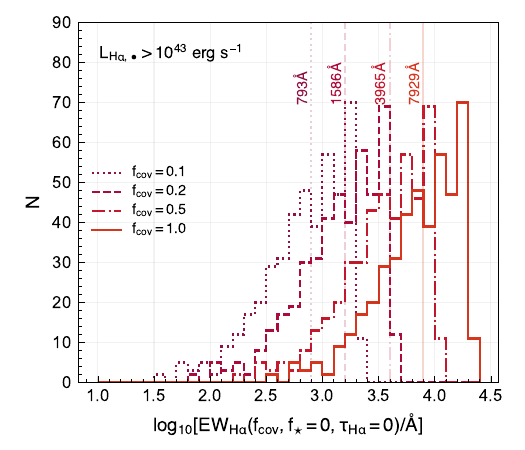}
    \includegraphics[width=1\columnwidth]{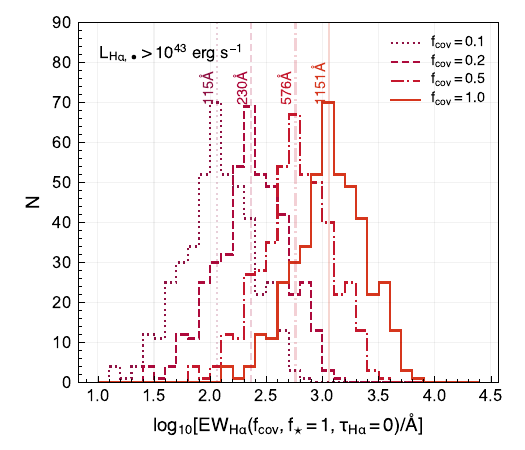}
    \caption{The distribution of BLAGN equivalent widths predicted for sources with \emph{intrinsic} $L_{\rm H\alpha, \bullet}/({\rm erg\ s^{-1}})>10^{43}$ for various assumed values of the covering fraction $f_{\rm cov}\in\{0.1, 0.2, 0.5, 1.0\}$. The top-panel shows the distribution ignoring the contribution of stars ($f_{\star}=0$) to the continuum while the bottom panel includes stellar continuum emission.}
    \label{fig:modelling:agn_ew_distribution}
\end{figure}

Next, in Figure \ref{fig:modelling:agn_ew_distribution},  we explore the equivalent width distribution predicted assuming a range of covering fractions $\fcov\in\{0.1, 0.2, 0.5, 1.0\}$ for sources with $L_{\rm H\alpha, \bullet}/{\rm erg\ s^{-1}}>10^{43}$. The top-panel of Figure \ref{fig:modelling:agn_ew_distribution} shows the predicted pure-AGN (i.e. $f_{\star}=0$) distribution. For $\fcov=0.5$ ($\fcov=0.2$) the median EW is approximately $4000{\rm\AA}$ ($1600{\rm\AA}$), both significantly in excess of the observational constraints ($\approx 500{\rm\AA}$). However, as noted, these distributions omit the contribution from stellar continuum emission. The bottom-panel of Figure \ref{fig:modelling:agn_ew_distribution} shows the result when the \emph{intrinsic} stellar emission is included. This sharply reduces the average (median) equivalent widths to $\sim 600{\rm\AA}$ ($250{\rm\AA}$) for $\fcov=0.5$ ($\fcov=0.2$), though the distribution is now broadened with many AGN having EWs $>1000{\rm\AA}$, assuming $\fcov=0.5$. Assuming that the disc, BLR, and stars are all equally affected by dust these EWs should remain unchanged. If the AGN is preferentially affected by dust, EWs would drop and vice versa.  

%% file: sections/conclusion.tex
\section{Discussion and limitations}\label{sec:discussion}

In the previous section we have demonstrated that it is possible to reconcile $z=5$ H$\alpha$ observations of BLAGN with predictions from the FLARES simulations combined with a relatively simple model. 

Despite our success there are however several limitations, which serve as the basis for future improvements:

In the preceding section, we have shown that observations of H$\alpha$ emission from BLAGN at $z=5$ can be brought into agreement with theoretical predictions from the FLARES simulations when coupled with a relatively simple forward-modelling framework. 

However, despite its success, our current model is subject to a number of important limitations. These constraints not only highlight areas where caution should be exercised in interpreting our results, but also point towards opportunities for future improvements. The following issues form the foundation for planned improvements in subsequent work:

\begin{itemize}

\item Although \flares\ simulates a larger effective volume than most other simulations of comparable resolution, it is evident that significant statistical uncertainties remain—particularly at the bright end of the luminosity function. Addressing this limitation will require extending the simulation to encompass larger volumes, which is a key priority for future work.

\item At present, observational samples at $z \geq 5$ remain limited — not only in terms of the number of confirmed sources, but also in the availability of complementary measurements such as line ratios and equivalent widths. Nevertheless, spectroscopic samples from JWST are expanding rapidly and are expected to be significantly augmented in the near future by data from Euclid.

\item To estimate the bolometric luminosity—and consequently the H$\alpha$ luminosity — we utilise supermassive black hole (SMBH) accretion rates averaged over a 10 Myr timescale, in order to suppress numerical noise inherent in the simulation. However, in reality, AGN exhibit variability on much shorter timescales, which are not captured by this approach. Thus, while we model the average, we expect tighter scatter than we would if we had modelled the variability. While it is not feasible to model such rapid variability self-consistently within the context of cosmological simulations, valuable insights can nonetheless be gained from observational studies and from high-resolution simulations of individual SMBH systems.

\item Our predictions are limited by the need to assume values for $\fblr$ and $\fcov$, the values of which are themselves sensitive to the averaging applied to the SMBH accretion rates. The uncertainty in these parameters gives us significant freedom to match observational constraints. However, these can potentially be constrained observationally, for example by conducting a complete survey for both Type I and Type II AGN and measuring additional quantities, including the equivalent widths and line ratios. 

\item While we have accounted for the impact of SMBH mass and accretion rate when calculating the disc SED, more realistic models \citep[e.g][]{Hagen2023} predict that the emission is also sensitive to the SMBH spin and disc orientation, something which most cosmological models do not currently predict \citep[though see e.g.][]{Husko2025, Beckmann2025}. 

\item The \qsosed\ model represents a simplified version of the more comprehensive \texttt{agnsed} framework, which has additional freedom in many of its modelling choices. 

\item In our modelling, we have assumed a uniform conversion efficiency of ionising photons into H$\alpha$ photons for all SMBHs. In reality, however, this conversion is highly sensitive to several factors, including the shape and intensity of the ionising radiation, as well as the physical conditions of the emitting gas—such as its geometry, density, chemical composition, and column density. As these properties can vary significantly between individual SMBHs, the actual conversion efficiency is expected to exhibit substantial source-to-source variation. To explore this further, we are undertaking a comprehensive set of photoionisation simulations aimed at quantifying how this conversion depends on these parameters, in addition to other key observables such as the Balmer decrement and break (Vijayan et al. \emph{in-prep}).

\item Although we have investigated the effects of dust attenuation, this analysis was not performed self-consistently using the same dust modelling framework applied to star particles within \flares. This decision was motivated by the fact that applying the standard model in the context of AGN-dominated galaxies results in extremely high levels of attenuation. Such extreme values may point to the necessity of incorporating an additional dust destruction mechanism operating in the immediate environments of AGN which requires higher resolution simulations to properly resolve.

\end{itemize}

\section{Conclusions}\label{sec:conclusions}

In this study, we have investigated the H$\alpha$ emission associated with actively accreting supermassive black holes (SMBHs) at high redshift ($z \geq 5$), as predicted by the First Light And Reionisation Epoch Simulations \citep[\flares][]{FLARES-I}. Our approach involved coupling the \flares\ simulations with the \qsosed\ accretion disc emission model to estimate the ionising photon output from each SMBH. We then applied a simplified photoionisation prescription to convert this ionising luminosity into predicted H$\alpha$ emission, assuming a fixed conversion efficiency.

The \qsosed\ model predicts that the shape of the SMBH accretion disc varies as a function of the BH mass and accretion rate. Consequently the SMBH ionising photon and H$\alpha$ bolometric corrections, and H$\alpha$ equivalent widths will vary these properties. To model the observational properties of AGN predicted by \flares\ we add additional complexity to our model: first, we assumed that only some fraction ($\fblr$) of AGN are observable as BLAGN; second, we assumed that only a fraction (the covering fraction, $\fcov$) of the disc emission is reprocessed by the broad-line emitting region (BLR); finally, we also consider the impact of dust attenuation relative to that predicted for the integrated stellar emission. This framework enables us to make predictions for the H$\alpha$ luminosity function and equivalent width distribution, which we then compare to recent observational constraints.

Comparison with observations favours a moderate, or larger ($>0.5$), value of $\fblr$ to reproduce observations of the faint-end slope of the luminosity function. Combined with $\fblr=0.75$ and no dust attenuation, a covering fraction of $\fcov\approx 0.2$ is able to simultaneously reproduce the observed faint and bright end of the BL AGN luminosity function. If instead AGN are attenuated similar to the stellar emission then a larger covering fraction ($\fcov\approx 0.5$) is required. Assuming $\fcov=0.2$ results in an average equivalent width $\approx 250{\rm\AA}$, lower than that observed ($\sim 500{\rm\AA}$). Assuming $\fcov=0.5$ instead produces average equivalent width comparable to the observational constraints. Thus, choosing $\fblr>0.5$, $\fcov\sim 0.5$, and assuming the AGN dust optical depth is similar to integrated stellar dust optical depth, we are then able reproduce current $z=5$ observational constraints, without any additional changes to the physics of the model.

Finally, while our framework is successful in reproducing current observations it is subject to several important limitations that we outline and discuss in detail. These include the requirement for simulations with larger effective volumes to better capture rare, high-luminosity objects; the need for more extensive and comprehensive observational datasets; the use of simplified photoionisation treatments that do not fully capture the diversity of AGN environments; and the current lack of a self-consistent approach to modelling dust attenuation in AGN-dominated systems.